\documentclass[twocolumn,showpacs,preprintnumbers,amsmath,amssymb,prl,runinaddre
ss,superscrip
taddress]{revtex4}

\usepackage{graphicx}
\usepackage{dcolumn}
\usepackage{bm}
\usepackage{tabularx}
\usepackage{amsmath}

\begin{document}

\title{Giant electron-electron scattering in the Fermi-liquid state of
 Na$_{0.7}$CoO$_2$}

\author{S. Y. Li}
\affiliation{D{\'e}partement de physique, Universit{\'e} de Sherbrooke,
 Sherbrooke, Qu{\'e}bec,
Canada}
\author{Louis Taillefer}
\affiliation{D{\'e}partement de physique, Universit{\'e} de Sherbrooke,
 Sherbrooke, Qu{\'e}bec,
Canada}
\author{D. G. Hawthorn}
\affiliation{Department of Physics, University of Toronto,
 Toronto, Ontario, Canada}
 \author{M. A. Tanatar}
 \altaffiliation{permanent address:  Institute of Surface Chemistry, N.A.S. Ukraine, Kyiv, Ukraine.}
 \affiliation{Department of Physics, University of Toronto,
 Toronto, Ontario, Canada}
 \author{Johnpierre Paglione}
\affiliation{Department of Physics, University of Toronto,
 Toronto, Ontario, Canada}
\author{M. Sutherland}
\affiliation{Department of Physics, University of Toronto,
 Toronto, Ontario, Canada}
\author{R. W. Hill}
\altaffiliation{now at: Department of Physics, University of Waterloo, Waterloo,
 Ontario, Canada}
\affiliation{Department of Physics, University of Toronto,
 Toronto, Ontario, Canada}
\author{C.H. Wang}
\affiliation{Structural Research Laboratory, University of Science and
 Technology of China,
Hefei, Anhui 230026, P. R. China}
\author{X.H. Chen}
\affiliation{Structural Research Laboratory, University of Science and
 Technology of China,
Hefei, Anhui 230026, P. R. China}

\date{\today}

\begin{abstract}
The in-plane resistivity, $\rho $, and thermal conductivity, $\kappa $, of
 single crystal
Na$_{0.7}$CoO$_2$
were measured down to 40 mK. Verification of the Wiedemann-Franz law,
$\kappa/T = L_0/\rho $ as $T \rightarrow 0$, and observation of a $T^2$
 dependence of $\rho $ at
low temperature establish the existence of a well-defined Fermi-liquid state.
 The measured
value of coefficient $A$ reveals enormous electron-electron scattering,
 characterized by the
largest Kadowaki-Woods ratio $A/\gamma^2$ encountered in any material.
The rapid suppression of $A$ with magnetic field
suggests a possible proximity to a magnetic quantum critical point. We also
 speculate on
the possible role of magnetic frustration and proximity to a Mott insulator.
\end{abstract}
\maketitle
Electron behavior in the layered cobaltate Na$_x$CoO$_2$ shows evidence of
 strong electron-electron
correlations: the specific heat at low temperature points to a significant mass
enhancement \cite{Ando}; the thermopower is ten times larger than that of
 typical metals at 300 K
\cite{Terasaki,Wang1}; the material can be made superconducting by intercalation
 with water \cite{
Takada}. These and other observations have stimulated extensive interest in this
 material and a
quest for possible new electronic phases \cite{Levi}.

A fundamental question is whether these strong electron correlations can be
 captured by the
standard model of metals, namely a Fermi-liquid (FL) description of the ground
 state and low-energy
excitations. A FL description is found to be generally valid for heavy-fermion
 materials, for
example, even though correlations in these systems lead to a huge
 renormalization of the electron
effective mass. On the other hand, such a description is {\it not} generally
 valid for cuprates,
except at the highest carrier concentrations. In this Letter, we report on two
 tests of FL theory
applied to Na$_x$CoO$_2$. The first is a test of the Wiedemann-Franz (WF) law,
 which determines
whether the delocalized fermionic excitations of the system carry charge $e$ and
 are therefore the
usual Landau quasiparticles. The second is a measurement of electrical
 resistivity at low
temperature, which looks at the lifetime of these quasiparticles and determines
 whether the
electron-electron scattering rate varies as $T^2$. We find that the WF law is
 satisfied and we
observe a clear $T^2$ regime in the resistivity:  $\Delta \rho = \rho - \rho_0 =
 AT^2$. What is
striking is the huge value of the coefficient $A$. When normalized by the
 quasiparticle effective
mass, it is two orders of magnitude larger than in heavy-fermion materials, such
 that the
Kadowaki-Woods ratio reaches an unprecedented value: $A/\gamma^2 \simeq 600~\mu
 \Omega$ cm mol$^2$
K$^2$/ J$^2$.

Na$_x$CoO$_2$ has a hexagonal layered structure consisting of stacked
 two-dimensional CoO$_2$
planes separated by spacer layers of Na$^+$ ions. The Co ions in each CoO$_2$
 plane are arranged on
a triangular lattice. In the undoped CoO$_2$ (without Na), each Co atom is in
 the Co$^{4+}$ valence
state with spin 1/2, and the material is speculated to be a Mott insulator. Due
 to the triangular
geometry, those spins are frustrated. With Na doping, each dopant atom
 contributes one electron,
changing Co$^{4+}$ to a spinless Co$^{3+}$ state. The effect of doping is to
 modify spin
correlations and introduce mobile charge carriers, just as in cuprates, but also
 relax magnetic
frustration, so that a rich interplay of spin and charge degrees of freedom is
 expected.

Single crystals of Na$_x$CoO$_2$ were grown from NaCl flux according to a
 procedure described
elsewhere \cite{Fujita}. The Na concentration was determined to be $x = 0.7$
 from a measurement of
the
$c$-axis lattice parameter (where $c$ = 10.94~\AA), using the calibration in
 Ref.~\onlinecite{Foo}.
Two samples, A and B, were cut to rectangular shapes with $\sim$mm dimensions in
 the $ab$ plane and 20-50 $\mu$m along the $c$ axis. 
Contacts were made with silver epoxy, diffused at 500 $^o$C for 1 hour, and were
 used to measure both electrical
resistivity  $\rho (T$) and thermal conductivity  $\kappa (T$) in a dilution
 refrigerator down to
40 mK. The
contact resistance was typically $\approx 0.1~\Omega $ at low temperature. The
 thermal conductivity
was measured
using a standard four-wire steady-state method with two RuO$_2$ chip
 thermometers calibrated
{\it in situ} against a reference Ge thermometer. Currents were made to flow in
 the $ab$ plane and
the
magnetic field was applied parallel to the current direction.

\begin{figure}[t]
\centering \resizebox{2.9in}{!}{\includegraphics{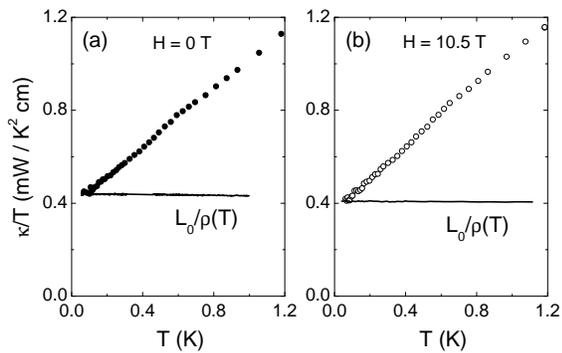}}
 \caption{\label{fig1}
Temperature dependence of the in-plane thermal conductivity, plotted as $\kappa
 (T)/T$, and the
in-plane electrical conductivity, plotted as $L_0/\rho (T)$, for sample A at two
 values of a magnetic field
applied parallel to the current: a) $H$ = 0, and b) 10.5~T. $L_0 = (\pi^2/3)
 (k_B/e)^2$. The
quantitative convergence of the two conductivities shows that the
 Wiedemann-Franz law is satisfied.
Note that the roughly linear increase in $\kappa (T)/T$ is due to phonon
 conduction.}
\end{figure}
In Fig.~1, we show the temperature dependence of the in-plane conductivity below
 1.2 K, both
electrical, plotted as $L_0/\rho (T)$ using the Lorenz number $L_0$ (see below),
 and thermal,
plotted as  $\kappa (T)/T$, in a magnetic field
$H$ = 0 and 10.5 T. There are two contributions to thermal conduction, coming
 respectively from
electrons and phonons. In the limit of electrons being scattered predominantly
 by defects (e.g. Na
impurities), the former will be linear in $T$. In the limit of phonons being
 scattered
predominantly
by electrons, the usual case for metals, phonon conduction is quadratic in $T$.
 This is indeed what
is observed in Na$_{0.7}$CoO$_2$ below 1~K or so: $\kappa = \alpha T + \beta
 T^2$. Irrespective of
the
particular temperature dependence, by taking the limit of $\kappa (T)/T$ as $T
 \rightarrow 0$ one
reliably obtains the purely electronic contribution at $T$ = 0. In the most
 general terms, this
residual linear term in the thermal conductivity, $\kappa_0/T$, is the entropy
 transport by the
delocalized fermionic excitations of the system. In a Fermi liquid, 
 $\kappa_0/T$ is directly
related to the residual electrical resistivity, $\rho_0$ = 55.8 $\mu \Omega$ cm
 for 0T,  via the Wiedemann-Franz 
(WF) law:
$\kappa_0/T = L_0/\rho_0$, where $L_0 \equiv (\pi^2/3) (k_B/e)^2 = 2.45 \times
 10^{-8}$~W~$\Omega$~
K$^{-2}$
is a universal constant. Fundamentally, the WF law says that Landau
 quasiparticles carry heat and
charge with strictly identical abilities, when no energy is lost through
 collisions. One
can see from Fig.~1 that the WF law is satisfied in Na$_{0.7}$CoO$_2$, both at
 $H$ = 0 and $H$ =
10.5~T.

In this particular sense, Na$_x$CoO$_2$ is like all other metals but one.
 Indeed, the WF law has
never been violated except in one instance: in the normal state of cuprate
 superconductor
Pr$_{2-x}$Ce$_x$CuO$_4$ at optimal doping ($x=0.15$) \cite{Hill}. However, at
 higher doping (in the
overdoped
regime), cuprates are found to also obey the WF law \cite{Proust,Nakamae}. This
 does not mean that
the electron behavior in Na$_x$CoO$_2$ necessarily conforms to FL theory. A
 basic property of FL
theory is that the electron-electron (or rather quasiparticle-quasiparticle)
 scattering rate should
grow as $T^2$. In the absence of any other inelastic scattering (such as
 electron-phonon), this
implies that the resistivity should have the form $\rho = \rho_0 +AT^2$. In the
 overdoped cuprate
Tl-2201 with a carrier concentration of 0.26 holes per Cu atom, the WF law is
 very accurately
obeyed but there is a prominent non-Fermi-liquid-like linear term in the
 resistivity at low
temperature: $\rho = \rho_0 + aT + bT^2$ \cite{Proust}. By going to the highest
 achievable
concentration of 0.3 holes per Cu atom in La$_{2-x}$Sr$_x$CuO$_4$, Nakamae {\it
 et al}. were able
to show
that cuprates do eventually show a purely quadratic dependence \cite{Nakamae}:
 $\rho = \rho_0
+AT^2$.
\begin{figure}[b]
\centering \resizebox{3.05in}{!}{\includegraphics{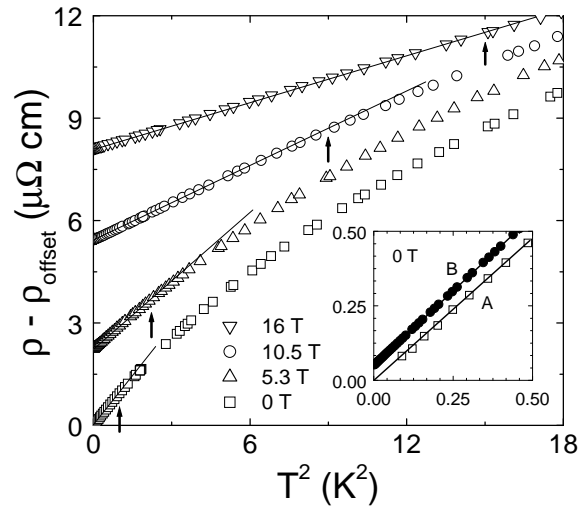}}
 \caption{\label{fig2}
Temperature dependence of the electrical resistivity at low temperature, plotted
 as
$\rho (T) - \rho_{\rm offset}$ vs. $T^2$, for sample A at several magnetic
 fields: $H$ = 0, 5.3,
10.5,
and 16 T.  $\rho_{\rm offset}$  is a constant arbitrary offset chosen for
 clarity of display. The
solid lines are
linear fits to the data in a range below a temperature $T_0$ indicated by
 arrows. The slope of
these
lines is the inelastic electron-electron scattering coefficient $A$, plotted vs.
 field in Fig.~3.
Inset: The zero-field data at the lowest temperatures.  Measurements on a second
 sample (B) show the $T^2$ 
behaviour down to 50 mK.}
\end{figure}

The in-plane resistivity $\rho(T)$ of Na$_{0.7}$CoO$_2$ at low temperature is
 shown in Fig.~2 for
different values of the magnetic field, plotted as $\rho - \rho_{\rm offset}$
 vs. $T^2$. A $T^2$
regime
is clearly observed for all fields, below a crossover temperature $T_0$ that
 grows with field. The
field dependence of $T_0$ and $A$ is shown in Fig.~3; $T_0$ goes roughly
 linearly from $T_0 = 1$~K
at $H=0$ to
$T_0 = 4$~K
at $H=16$~T while $A$ decreases from 0.96 to 0.22 $\mu \Omega$ cm K$^-2$. (The
 low
value of $T_0$ explains why the $T^2$ dependence had not been seen in
previous studies \cite{Bruhwiler,Rivadulla}.) The observed relation $\Delta \rho
 = AT^2$ indicates
that the behavior of electrons in this system is well described by FL theory.
 The remarkable aspect
is that the magnitude of the electron-electron scattering is enormous: in zero
 field, $A$ = 1.0
$\mu \Omega$ cm K$^{-2}$.  This is as large as in heavy-fermion systems, where
 the strong
quasiparticle-quasiparticle scattering is due to the enormous density of states
 at the Fermi
energy, as measured by the residual linear term in the specific heat, $\gamma =
 C/T$ as
$T\rightarrow 0$, or equivalently the huge effective mass, $m^*$, as measured
 for example by the de
Haas-van Alphen effect. In these materials, $A$ is found to be roughly
 proportional
to $\gamma^2$. In UPt$_3$, for
example, the
relation $A \propto \gamma^2$ holds very well as a function of pressure
 \cite{Joynt}. In fact quite
generally, the ratio $r_{KW} \equiv A/\gamma^2$, known as the Kadowaki-Woods
 ratio, has been shown
to
have a nearly universal value of about 10 $\mu \Omega$ cm mol$^2$ K$^2$ / J$^2
 \equiv a_0$ in
heavy-fermion systems \cite{
Kadowaki, Miyake}
(where $\gamma$ is measured per mole of magnetic ion).
\begin{figure}[t]
\centering \resizebox{2.65in}{!}{\includegraphics{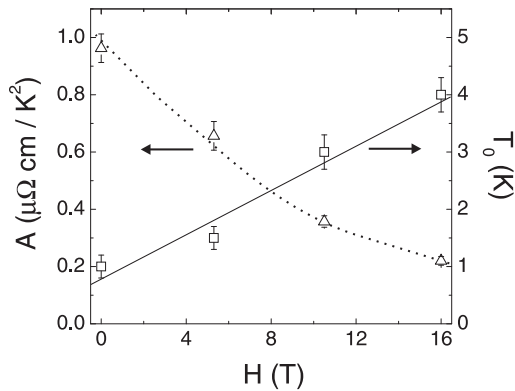}}
 \caption{\label{fig3} Field
dependence of the $T^2$ coefficient $A$ (in $\Delta \rho = AT^2$), and the upper
 limit of the $T^2$
range, $T_0$ (arrows in Fig.~2), for sample A. Lines are guides to the eye.}
\end{figure}
\begin{figure}[b]
\centering \resizebox{2.75in}{!}{\includegraphics{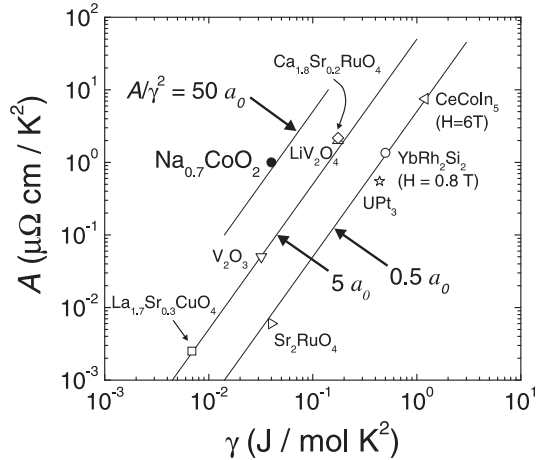}}
 \caption{\label{fig4}
Kadowaki-Woods plot of coefficient $A$ (in $\Delta \rho = AT^2$) vs. $\gamma$ ,
 the residual linear
term in the specific heat  ($\gamma = C/T$ as $T \rightarrow 0$), for a number
 of metals. The three
lines are lines of constant Kadowaki-Woods ratio $r_{KW} = A /\gamma^2$, for
 values of 0.5, 5
and 50 $a_0$, as indicated. The first line at 0.5 $a_0$ is characteristic
of
heavy-fermion materials and also accounts for the quasi-2D Fermi liquid
 Sr$_2$RuO$_4$; the second
line at 5 $a_0$ corresponds to the highest values observed until now (typically
 in systems with
magnetic frustration or close to a Mott insulator); the third line at 50 $a_0$
 shows the
order-of-magnitude larger value found in Na$_{0.7}$CoO$_2$.
(The data used in this plot are referenced in the text.)  The data for
 YbRh$_2$Si$_2$ is for a
sample doped with 5\% Ge, for which the QCP is pushed to very low fields (see
 text).}
\end{figure}

In Fig.~4, a log-log (``Kadowaki-Woods'') plot of $A$ vs. $\gamma$ is reproduced
 for a number of
materials. In such a plot, it is important to take into account the effect of
 anisotropy, since $A$
is in general dependent on current direction whereas $\gamma$ is an average over
 all directions on
the Fermi surface. In the 3D metal UPt$_3$, for example, that hexagonal crystal
 structure leads to
a mass tensor anisotropy of 2.7 which is reflected in the conductivity of the FL
 regime (below
$T_0 \simeq$ 1.5 K), where one finds $A$ = 0.55 (1.55) $\mu \Omega$ cm K$^{-2}$
 for a current
parallel (perpendicular) to the hexagonal $c$-axis \cite{Joynt}. In Fig.~4, we
 use the lower value
(direction of maximum conductivity) and  $\gamma$ = 0.44 J / mol K$^2$
 \cite{Joynt}, so that
$r_{KW}$
= 0.28 $a_0$. In quasi-2D systems, transport anisotropy can be much
more
extreme. The best characterized example of a Fermi liquid state with quasi-2D
 conductivity is the
layered ruthenate Sr$_2$RuO$_4$ \cite{Maeno}. Conduction perpendicular to the
 RuO$_2$ planes is
only
coherent at low temperature, and the mass tensor anisotropy of the Fermi surface
 is 1000 or so.
This
gives rise to a FL regime below $T_0 \simeq$ 20 K with $A \simeq$ 0.006 (6) $\mu
 \Omega$ cm
K$^{-2}$,
for in-plane (out-of-plane) transport \cite{Maeno}. Given that the quasiparticle
 states essentially
all have wavevectors in the plane, it only makes sense to compute a
 Kadowaki-Woods ratio for
in-plane transport: with  $\gamma $ = 0.04 J / mol K$^2$ \cite{Maeno}, we get
 $r_{KW} \simeq$
0.4 $a_0$.

The striking fact about the quasi-2D cobalt oxide Na$_x$CoO$_2$ is that although
 it has the same
$\gamma$  as the quasi-2D ruthenium oxide, its in-plane $A$ coefficient is two
 orders of magnitude
larger, {\it i.e.} of the same magnitude as in heavy-fermion systems.
Indeed in Na$_{0.7}$CoO$_2$, $\gamma$ = 0.04 J / mol K$^2$ (per mole of Co)
\cite{Bruhwiler},
so that the Kadowaki-Woods ratio is more than 100 times larger than in
 Sr$_2$RuO$_4$:
$r_{KW} \simeq 60~a_0$! This reveals that
the strong electron correlations in Na$_x$CoO$_2$ are responsible for enhancing
 not so much the
effective mass of quasiparticles as their scattering rate. Enhanced values of
 $r_{KW}$ beyond the
typical value of 10 $\mu \Omega$ cm mol$^2$ K$^2$ / J$^2$ have been seen in a
 few cases. We now
consider these examples to suggest possible mechanisms for the huge $r_{KW}$
 value in
Na$_{0.7}$CoO$_2$.

The first possible mechanism is proximity to a quantum critical point (QCP). In
 the
heavy-fermion material YbRh$_2$Si$_2$, a magnetic-field-induced QCP occurs when
antiferromagnetic order is suppressed by applying a field greater than a
 critical field $H_c$
\cite{Custers}. This
leads to a divergence of both $A$ and $\gamma $  as $H \rightarrow H_c$, where
$A \sim (H-H_c)^{-\alpha}$, with power $\alpha $ = 1.0. The value of $H_c$ can
 be made very small (30
mT) by substituting 5\% of Si for Ge. In YbRh$_2$(Si$_{0.95}$Ge$_{0.05}$)$_2$
 \cite{Custers},
$r_{KW}$ is roughly independent of field at large values of the field: it is
 constant at 0.54
$a_0$ for $(H-H_c)/H_c >$ 10. However, as the field is lowered
towards
$H_c$, a distinct rise in $r_{KW}$ is observed, reaching a value of
2 $a_0$ at
$(H-H_c)/H_c \simeq $ 1. A similar field-tuned QCP is observed in the
 heavy-fermion material
CeCoIn$_5$ \cite{Paglione,Bianchi}, with $H_c$ = 5.1 T and $\alpha $ = 4/3,
 where one finds $A$ =
7.5 (1.0) $\mu \Omega $ cm / K$^2$~\cite{Paglione} and  $\gamma $ = 1.2 (0.64) J
 / mol K$^2$~
\cite{Bianchi}, so that $r_{KW}$ = 0.52 (0.24) $a_0$ at $H$ = 6 (9) T.
So here
again a field-induced enhancement of $r_{KW}$ is observed as one approaches the
 QCP. It is
interesting to note that a similar effect is observed in Na$_{0.7}$CoO$_2$. In
 Fig.~3, the
coefficient
$A$ determined from data in Fig.~2 is plotted as a function of magnetic field.
 While there is no
divergence {\it per se}, a five-fold increase is nevertheless observed as $H$
 goes from 16 to 0 T.
Using
the specific heat data of Br{\"u}hwiler {\it et al}. \cite{Bruhwiler},
where $\gamma $  = 0.04 (0.025) J / mol K$^2$ in $H$ = 0 (14) T,
we get $r_{KW}$ = 60 (40) $a_0$, at $H$ = 0
(14) T. This therefore suggests that one interpretation of the strong field
 dependence of $A$ is a
close-by QCP of magnetic nature. In support of this interpretation, there is
 evidence of a spin
density wave transition in Na$_x$CoO$_2$ at a slightly higher Na concentration,
 namely $x$ = 3/4
\cite{Motohashi,Sugiyama,Sugiyama2}, and  $\rho (T$) below 20 K is indeed
 steepest at
$x$ = 0.75 \cite{Foo},
while $\gamma$ remains constant between $x$ = 0.55 \cite{Ando} and 0.7
 \cite{Bruhwiler}.  Quantum critical behaviour has also been theoretically
 predicted in Na$_x$CoO$_2$ based on discrepancies between ferromagnetism
 predicted by density-functional calculations and the measured paramagnetic
 ground state \cite{Singh}.

A second comparison, to the transition metal oxide LiV$_2$O$_4$, is highly
 suggestive. Both the
cubic spinel structure of LiV$_2$O$_4$, in which V ions lie on a sub-lattice of
 corner-sharing
tetrahedra, and the layered structure of Na$_x$CoO$_2$, in which Co ions lie on
 a two-dimensional
triangular lattice, give rise to strong magnetic frustration. The
 extraordinarily heavy mass
observed in the FL state of LiV$_2$O$_4$ - characterized by $\gamma \simeq$
 0.175 J / mol K$^2$
(per mole
of V), $A$ = 2.0 $\mu \Omega $ cm / K$^2$ and $T_0 \simeq$ 1.5 K \cite{Urano} -
 has been
attributed to geometric frustration \cite{Urano,Lacroix}. Although this has not
 been emphasized,
this material has also held the record for the largest Kadowaki-Woods ratio
 until now: $r_{KW}$ =
6.5 $a_0$. This is one order of magnitude larger than in typical
heavy-
fermion materials - like YbRh$_2$Si$_2$ (0.54), UPt$_3$ (0.28) and CeCoIn$_5$
 (0.24) - but still
one order of magnitude {\it smaller} than in Na$_x$CoO$_2$. The latter
 discrepancy might have to
do
with the different dimensionalities of the electron system in these two oxides:
 3D in LiV$_2$O$_4$,
2D in Na$_x$CoO$_2$.

Finally, we mention a third instance of anomalously large Kadowaki-Woods ratios:
proximity to a Mott insulator.  The transition metal oxide V$_2$O$_3$ is close
 to a metal-insulator
transition and it has $\gamma \simeq$ 0.032 J / mol K$^2$ (per mole of V) and
 $A$ = 0.05
$\mu \Omega$ cm / K$^2$, so that $r_{KW} \cong$ 5 $a_0$ \cite{Miyake}.
Replacing Sr by
Ca in Sr$_2$RuO$_4$ produces a Mott insulator. By gradually replacing Sr,
 Nakatsuji {\it et al}.
have
shown how
$r_{KW}$ in Ca$_{2-x}$Sr$_x$RuO$_4$ goes from 0.4 $a_0$ at $x$ = 2, to $a_0$ at
 $x$ = 0.5, to a
value as large
as
7 $a_0$ at $x$ = 0.2 \cite{Nakatsuji}.
For $x < 0.5$, a structural transition appears to induce a Mott gap on part of
 the Fermi surface,
with
concomitant $S$ = 1/2 localized moments, as evidenced by a decrease of $\gamma$
 with decreasing
$x$. The authors
interpret the large enhancement of $A$ near the Mott insulator as a combination
 of the narrowing of
the remaining conduction band and the additional scattering of conduction
 electrons on these
(antiferromagnetically-coupled) localized moments \cite{Nakatsuji}. In this
 respect, we note that
cuprates, the most infamous
doped Mott insulators, also exhibit anomalously large $r_{KW}$. Specifically, in
 overdoped
La$_{2-x}$Sr$_x$CuO$_4$
with $x$ = 0.3, the FL state is characterized by $\gamma \simeq$ 7 mJ / mol
 K$^2$, $A \simeq$
2.5 $\mathrm{n}\Omega$ cm / K$^2$ and $T_0 \simeq $ 1.5 K
\cite{Nakamae}, so that $r_{KW} \simeq 5~a_0$. Although this
doping
value ($x$ = 0.3) is not usually thought
to be close to the Mott insulator ($x$ = 0), $r_{KW}$ is still 10 times larger
 than in the
isostructural
material Sr$_2$RuO$_4$. (Note that in none of these cases is geometric
 frustration an issue.)

In conclusion, Na$_x$CoO$_2$ with $x$ = 0.7 adopts a Fermi-liquid state at low
 temperature that is
characterized by the largest Kadowaki-Woods ratio ever observed. Comparison with
 other materials
suggests that the unprecedented magnitude of the electron-electron scattering is
 due either to
magnetic frustration or to the proximity of a nearby magnetic quantum critical
 point or a Mott
insulator. Theoretical exploration of the impact of magnetic frustration on
 electron scattering,
including the role of magnetic field and reduced dimensionality, would be most
 useful. 

This work was supported by the Canadian Institute for Advanced Research and a
 Canada Research Chair, and funded by NSERC of Canada.

\end{document}